\documentclass[twocolumn]{autart}
\usepackage{silence}
\usepackage{algpseudocode}
\usepackage{graphicx}
\usepackage{amsmath, amsfonts}
\usepackage[colorlinks,citecolor=blue,linkcolor=blue]{hyperref}
\usepackage[round,longnamesfirst]{natbib}
\usepackage{bm}
\usepackage{mathrsfs}
\usepackage{changepage}
\usepackage{enumitem}
\usepackage{xcolor}
\usepackage{soul}
\usepackage{tikz}
\usepackage{float}
\usepackage{subcaption}
\usepackage[T1]{fontenc}
\usepackage[skip=0pt]{caption}
\usepackage{csquotes}
\graphicspath{{./figures/}}
\DeclareMathOperator{\interior}{int}
\usepackage[normalem]{ulem}

\usepackage{etoolbox}

\makeatletter
\def\@xnamedef#1{\expandafter\protected@xdef\csname #1\endcsname}
\def\no@harm{} 
\def\ead@au#1{\protected@edef\@ead@au{#1}}
\patchcmd\runningauthor@fmt{\global\edef}{\protected@xdef}{}{}
\patchcmd\runningauthor@fmt{\global\edef}{\protected@xdef}{}{}
\patchcmd\author@fmt{\edef}{\protected@edef}{}{}
\patchcmd\add@xtok{\xdef}{\protected@xdef}{}{}
\makeatother
\renewcommand{\j}{\textrm{\normalfont{j}}}

\begin{document}
    \begin{frontmatter}
        \runtitle{Stability analysis of time-delay systems in the parametric space}
        \title{Stability analysis of time-delay systems in the parametric space\thanksref{footnoteinfo}}

        \thanks[footnoteinfo]{This paper was not presented at any IFAC meeting.
                              Corresponding author Vukan Turkulov. Tel. +381621829124}

        \author[NoviSad]{Vukan Turkulov}\ead{vukan\_turkulov@uns.ac.rs},
        \author[NoviSad]{Milan R. Rapai\'{c}}\ead{rapaja@uns.ac.rs},
        \author[Bordeaux]{Rachid Malti}\ead{firstname.lastname@ims-bordeaux.fr}

        \address[NoviSad]{University of Novi Sad, Faculty of Technical Sciences, Trg Dositeja Obradovi\'{c}a 6, Serbia }
        \address[Bordeaux]{Univ. Bordeaux, CNRS, Bordeaux INP, IMS, UMR 5218, F-33400 Talence, France}


        \begin{abstract}
            This paper presents a novel method for stability analysis of a wide class of linear,
            time-delay systems (TDS), including retarded, incommensurate and distributed delays.
            The proposed method is based on frequency domain analysis and application of Rouch\'e's theorem.
            Given a parametrized TDS and an arbitrary parametric point,
            the proposed method is capable of identifying the surrounding region in the parametric space
            for which the number of unstable poles remains invariant.
            First, a procedure for investigating stability along a line is developed.
            Then, the results are extended by application of
            H\"older's inequality to investigate stability within a region.
            The proposed method is uniformly applicable to parameters of
            different types (simple delays, distributed delay limits, time constants, etc.),
            as illustrated by examples.
        \end{abstract}
    \end{frontmatter}

    \section{Introduction}
    \label{section:introduction}
    Time delays are effectively used to model a wide range of physical, economic, social and biological phenomena.
    Examples include modeling industrial processes and their control,
    epidemic dynamics, operations research and computer network flows.
    TDS are infinite-dimensional,
    rendering their behavioral analysis more challenging as compared to their finite-dimensional counterparts.

    The methodology presented in this paper performs stability analysis in a given parametric space.
    Thus, it is natural to compare it to $\mathcal{D}$-partitioning
    methods (\cite{neimark1949, gryazina2004, neimark1998, lee1969, elsgolts1973}).
    Such methods view the parametric space as being split into multiple partitions,
    with an invariant number of unstable poles inside each individual partition.
    In that context, the proposed method determines one such partition,
    starting from any of its interior points.
    The method finds the entire partition, regardless of its shape.
    The stability can be investigated with respect to both delays and other types of parameters.

    Similarities can also be drawn towards methods which determine the parametric stability crossing set (SCS).
    The SCS is defined as the collection of surfaces in the parametric space
    for which there is at least one system pole on the imaginary axis.
    Such approaches have been successfully developed for retarded systems with two and three independent delays
    (\cite{hale1993,gu2005,sipahi2005,gu2011}),
    providing insightful graphical representation of stability equivalence regions.
    Similar methods have been proposed in the domain of robust control (\cite{morarescu2006}).
    Alternatively, instead of computing the SCS in a high dimensional parametric space,
    it is possible to directly compute the projection of SCS to a
    low dimensional space (\cite{sipahi2009,delice2010}).
    Finally, SCS-based methods may be used to determine the stability radius
    of a given parametric point (\cite{gu2007}).
    The method proposed in the present paper bears similarities with frequency sweeping stability analysis methods,
    such as the ones proposed in \citep{chen1995, niculescu1999, li2013, li2015, li2017}.
    The nature of similarities is technical, as the proposed method involves frequency sweeping tests.
    On the other hand, the proposed method differs from the aforementioned ones in terms of problem formulation,
    classes of applicable systems and/or the resulting conservatism.
    The stability boundary in the parametric space can also be found by approximating an infinite-dimensional
    system with a finite one, as proposed in \cite{breda2009}.
    The method proposed in this work uses no such approximations.

    The methodology proposed in this paper is also applicable to systems containing distributed delays.
    Stability analysis of such systems is challenging due to their form,
    which is less well-behaved compared to their discrete delay counterparts.
    Interesting techniques for stability analysis of such systems can be found in \cite{morarescu2007,gu2003,zeng2015}.
    General behavioral analysis of TDS can be found in papers such as
    \cite{datko1978,cooke1982,bellman1963,michiels2007}.
    An overview of existing TDS stability analysis methods is provided in several books,
    including \cite{dugard1998,gu2003,niculescu2004,wu2010,fridman2014,michiels2014}.

    The strengths of the presented method are summarized as follows.
    Firstly, it allows determining whether two parametric points
    have the same stability characteristics with an algorithmic complexity
    independent of the number of parameters,
    when both points belong to the same convex stability equivalence region.
    Secondly, the entire stability equivalence region is determined without any conservatism.
    It is worth emphasizing that the method is applicable to a 
    broad class of linear TDS,
    including retarded, incommensurate and distributed delay systems.
    A simplified methodology, giving stability conditions along a parametric line in the case of
    a specific system involving two delays, was previously considered in \cite{turkulov2019}.

    \paragraph*{Notations.}
    \label{section:notation}
    The paper utilizes standard mathematical notations.
    Symbol $s$ denotes the Laplace variable.
    Angled brackets $\langle\cdot,\cdot\rangle$ represent the dot product.
    The $p$-norm of a vector $\mathbf{x}$ is denoted as $||\mathbf{x}||_p$.
    The set of non-negative real numbers is denoted as $\mathbb{R}_0^+$
    and the set of non-negative integers by $\mathbb{N}_0$.
    Boundary of set $\mathcal{X}$ is denoted $\partial \mathcal{X}$
    and the interior of set $\mathcal{X}$ is denoted $\interior \big(\mathcal{X}\big)$.
    The expressions ''left-hand side'' and ''right-hand side'' are abbreviated to LHS and RHS, respectively.
    The Bromwich-Wagner contour enveloping the entire right half of the complex plane
    is denoted as $\mathcal{C}$ and defined as
    \begin{equation}
        \begin{split}
            \mathcal{C} &= \mathcal{C}_a \cup \mathcal{C}_c \\
            \mathcal{C}_a &= \{s = \j\omega \; | \; \omega \in \mathbb{R}\} \\
            \mathcal{C}_c &= \left\{s = \lim_{\rho \rightarrow \infty} \rho e^{\j\varphi} \;
                \Big| \; \varphi \in \left(-\frac{\pi}{2}, \frac{\pi}{2}\right)\right\}
        \end{split}
    \end{equation}
    The characteristic function of a linear TDS is defined as
    \begin{equation}
        f:\mathbb{C} \times \mathcal{T} \rightarrow \mathbb{C},
        \label{eq:characteristic_definition}
    \end{equation}
    where $\mathcal{T} \subset (\mathbb{R}_0^+)^n$ denotes a parametric space.
    A parametric point is denoted as $\bm{\tau} = [\tau_1, \tau_2, \dotsc, \tau_n] \in \mathcal{T}$.
    The gradient vector field of $f$ over the parametric space is denoted as $\nabla f$.
    $NU_f(\bm{\tau})$ designates the number of zeros of the characteristic function $f(s, \bm{\tau})$
    with non-negative real part, where each zero is counted as many times as its multiplicity.
    The set of all parametric points of $f(s, \bm{\tau})$ sharing the same number of zeros
    with non-negative real part as a starting point $\bm{\tau^0} \in \mathcal{T}$ is defined as
    \begin{equation}
        \mathcal{M}_{f}^{\#}(\bm{\tau^0}) =
            \{ \bm{\tau} \in \mathcal{T} \quad | \quad NU_f(\bm{\tau}) = NU_f(\bm{\tau^0}) \}.
        \label{eq:mf}
    \end{equation}
    Define the maximum surrounding stability equivalence region of $\bm{\tau^0}$, $\mathcal{M}_f(\bm{\tau^0})$,
    as a set of points $\bm{\tau}$ satisfying the following conditions:
    \begin{enumerate}
        \item $\bm{\tau} \in \mathcal{M}_{f}^{\#}(\bm{\tau^0}) \subset \mathcal{T}$
        \item There exists a path $\mathcal{P}$ which connects $\bm{\tau^0}$ with $\bm{\tau}$,
              such that $\mathcal{P} \subset \interior \big(\mathcal{M}_f^{\#}(\bm{\tau^0})\big)$.
    \end{enumerate}

    Define \textquote{stability equivalence segment (or region)} of $f$
    as the segment (or region) that has an equivalent number of unstable poles,
    i.e. for which $NU_f(\bm{\tau})$ is invariant.  
    When $NU_f(\bm{\tau}) = 0$,
    it designates a stability segment (or region).
    When $NU_f(\bm{\tau}) > 0$, it designates an instability segment (or region) having the same number of unstable poles.

    \paragraph*{Paper outline.}
    The paper is organized as follows:
    section \ref{section:problem} defines problems considered in the remainder of the paper.
    The main results of the paper are presented in sections \ref{section:line} and \ref{section:region}.
    Section \ref{section:line} lays out the theory for extending the stability along a line,
    with additional adaptations well-suited for retarded TDS.
    Section \ref{section:region} extends the methodology to analyze stability within a region.
    Methods presented in sections \ref{section:line} and \ref{section:region}
    are illustrated on examples with retarded and non-retarded TDS.
    Finally, section \ref{section:conclusion} presents a short summary with several closing comments.

    \section{Problem definition}
    \label{section:problem}
    Consider a linear TDS with a characteristic function $f(s, \bm{\tau})$ given in an explicit form.
    Starting from a parametric point $\bm{\tau^0} \in \mathcal{T}$, two versions of the problem are defined:
    \begin{enumerate}[label=(P\arabic*)] 
        \item \label{problem1} \textbf{Stability equivalent segment}.
              Find the maximum segment $\mathcal{E} \subset \mathcal{T}$ along a
              predefined direction originating from $\bm{\tau^0}$ such that
              $NU_f(\bm{\tau^0}) = NU_f(\bm{\tau}), \forall \bm{\tau} \in \mathcal{E}$.
        \item \label{problem2} \textbf{Stability equivalence region}. Find the maximum stability region       
        $\mathcal{M}_{f}(\bm{\tau^0})$, surrounding $\bm{\tau^0}$.
    \end{enumerate}

    Likewise, the paper presents two versions of the method (line-based in section \ref{section:line}
    and region-based in section \ref{section:region}) for solving both problems.
    For the method to be applicable, the following hypotheses must hold:
    \begin{enumerate}[label=(H\arabic*)]
        \item \label{hypothesis1}
            System characteristic function must be holomorphic in the open right half complex-plane,
            continuous on the imaginary axis for all $\bm{\tau} \in \mathcal{T}$
            and continuously differentiable with respect to $\bm{\tau}$ in the closed right half complex-plane.
            These conditions hold for a majority of TDS,
            but they fail for most systems with spatially distributed and/or fractional dynamics.
        \item \label{hypothesis2}
            The characteristic function must satisfy
            \begin{equation}
                \lim_{\rho \rightarrow \infty} \frac{\big| f(\rho e^{\j \varphi}, \bm{\tau^A})\big|}
                {\left|\int_{\bm{\tau^A}}^{\bm{\tau^B}}
                    \langle \nabla f(\rho e^{\j\varphi}, \bm{\tau}), d\bm{\tau}\rangle
                \right|} = \infty,
                \label{eq:h2}
            \end{equation}
            $\forall \bm{\tau^A}, \bm{\tau^B} \in \mathcal{T}$,
            $\forall \varphi \in \left[-\frac{\pi}{2}, \frac{\pi}{2}\right]$,
            where $\int_{\bm{\tau^A}}^{\bm{\tau^B}}$ denotes a line integral along a curve $\gamma$
            connecting points $\bm{\tau^A}$ and $\bm{\tau^B}$ such that $\gamma \subset \mathcal{T}$.
    \end{enumerate}
    The hypotheses (H1) and (H2) are the only conditions for the results of this paper to hold.
    These hypotheses are satisfied by a wide class of systems,
    including all retarded and some distributed delay systems.
    For example, it can easily be proven that (H1) and (H2) hold for all characteristic functions of the form
    \begin{equation}
        f(s, \bm{\tau}) = s^m + \sum_{i=0}^{m-1} s^i
        \left( \sum_{k=1}^n \alpha_{i,k}(\bm{\tau}) e^{-s \beta_{i,k}(\bm{\tau})} \right),
        \label{eq:wide_retarded_systems}
    \end{equation}
    where $\alpha_{i,k}(\bm{\tau}), \beta_{i,k}(\bm{\tau}) : \mathcal{T} \rightarrow \mathbb{R}$
    are differentiable functions for $i = 0, 1, \cdots, m-1$, $k = 1, 2, \cdots, n$ and
    $\beta_{i,k}(\bm{\tau}) \geq 0, \forall \bm{\tau} \in \mathcal{T}$.

    Although the results, presented in sections \ref{section:line} and \ref{section:region},
    are valid for all kind of TDS satisfying (H1) and (H2),
    special attention is given to TDS of retarded type
    as they introduce further simplifications to the established results.
    Finally, it is important to stress that the stability addressed in this paper is of exponential type.
    A similar method, investigating BIBO stability of fractional non-commensurate systems subject
    to perturbations in differentiation orders, is proposed in \cite{rapaic2019}.

    \section{Stability equivalence along a line}
    \label{section:line}
    In this section, a solution to problem \ref{problem1} is obtained.
    Let us characterize variations of $\bm{\tau}$ along a line starting from $\bm{\tau^0}$
    by a single scalar non-negative parameter $\theta$ as
    \begin{equation}
        \bm{\tau} (\theta) = \bm{\tau^0} + \theta \bm{\tau^d}, \qquad \theta \geq 0
        \label{eq:line_parametrization}
    \end{equation}
    where $\bm{\tau^d}$ is an arbitrarily chosen unit direction vector.
    Define the starting value of $\theta$ as $\theta_0 = 0$, corresponding to $\bm{\tau}(0) = \bm{\tau^0}$.
    For simplicity, in this section, the characteristic function
    is expressed as $f(s, \bm{\tau}(\theta)) \equiv f(s, \theta)$.
    The Problem \ref{problem1} reduces to finding the maximum value of $\theta$
    for which the number of non-negative zeros of $f$ is preserved.
    Such stability-limiting value of $\theta$ is defined as
    \begin{equation}
        \theta_{lim} = \sup {
            \left\{\theta^* \Big| NU_f(\bm{\tau}(\theta))=NU_f(\bm{\tau_0})
                \;, \; \forall \theta \in [\theta_0, \theta^*)
            \right\}} \;.
        \label{eq:theta_lim}
    \end{equation}

    \subsection{Sufficient condition}
    As a first step towards finding $\theta_{lim}$, sufficient stability equivalence condition along a line is provided.
    \begin{thm}
        Let $f$ satisfy hypotheses \ref{hypothesis1} and \ref{hypothesis2}.
        Let $\theta_0 \geq 0$ be an initial point such that
        $f(\j\omega, \bm{\tau}(\theta_0)) \neq 0, \forall \omega \in \mathbb{R}$.
        Let $\bm{\tau}(\theta)$ be defined as in \eqref{eq:line_parametrization}.
        Then,
        \begin{equation*}
            NU_f(\bm{\tau}(\theta_0)) = NU_f(\bm{\tau}(\theta_0 + \Delta))
        \end{equation*}
        holds for all $0 \le \Delta < \overline{\Delta}(\theta_0)$, where
        \begin{equation}
            \overline{\Delta}(\theta_0) = \min_{\omega \in \mathbb{R}_0^+}\frac{|f(\j\omega, \theta_0)|}
            {\displaystyle \max_{\theta_0 \leq \beta \leq \theta_0 + \overline{\Delta}(\theta_0)}
            \Big|\frac{\partial f}{\partial \theta} (\j\omega, \theta=\beta) \Big|}.
            \label{eq:line_sufficient}
        \end{equation}
        \label{thm:line_sufficient}
    \end{thm}
    \vspace{-0.7cm}
    \begin{pf}
        Due to \ref{hypothesis1}, Rouch\'e's theorem  can be applied to $f$ as
        \begin{multline}
            |f(s,\theta_0 + \Delta) - f(s, \theta_0)| < |f(s, \theta_0)|,
            \forall s \in \mathcal{C} \Rightarrow\\ NU_f(\bm{\tau}(\theta_0)) = NU_f(\bm{\tau}(\theta_0 + \Delta)).
            \label{eq:rouche}
        \end{multline}
        Furthermore, the fundamental theorem of calculus can be applied to the inequality in \eqref{eq:rouche},
        resulting in
        \begin{equation}
            \Bigg| \int_{\theta_0}^{\theta_0 + \Delta}
            \frac{\partial f}{\partial \theta} (s, \theta=\beta)d\beta \Bigg|< |f(s, \theta_0)|.
            \label{eq:fundamental}
        \end{equation}
        Due to \ref{hypothesis2}, inequality \eqref{eq:fundamental} holds $\forall s \in \mathcal{C}_c$.
        Taking the symmetry of $f(s, \theta)$ into account,
        further analysis is restricted to $s = \j\omega, \forall \omega \in \mathbb{R}_0^+$.
        Notice that
        \begin{multline}
            \Bigg|\int_{\theta_0}^{\theta_0 + \Delta}
            \frac{\partial f}{\partial \theta} (\j\omega, \theta=\beta) d\beta\Bigg| \leq\\
            \int_{\theta_0}^{\theta_0 + \Delta}\Bigg|
            \frac{\partial f}{\partial \theta} (\j\omega, \theta=\beta) \Bigg| d\beta \leq \\
            \Delta \max_{\theta_0 \leq \beta \leq \theta_0+\Delta}
            \Big|\frac{\partial f}{\partial \theta} (\j\omega, \theta=\beta) \Big|.
            \label{eq:conservative_integral}
        \end{multline}
        Introducing the conservative bound \eqref{eq:conservative_integral} into \eqref{eq:fundamental} yields
        \begin{equation}
            \Delta \cdot \max_{\theta_0 \leq \beta \leq \theta_0+\Delta}
            \Big|\frac{\partial f}{\partial \theta} (\j\omega, \theta=\beta) \Big|  < |f(\j\omega, \theta_0)|.
            \label{eq:fundamental2}
        \end{equation}
        The LHS of \eqref{eq:fundamental2} is non-decreasing, as a product of two non-decreasing functions.
        Consequently, if inequality \eqref{eq:fundamental2} holds for some value of $(\theta_0 + \Delta)$,
        it also holds for all values of $\beta \in [\theta_0, \theta_0+\Delta]$.
        Based on this fact,
        \eqref{eq:fundamental2} yields
        \begin{equation}
            \Delta < \frac{|f(\j\omega, \theta_0)|}  {\displaystyle \max_{\theta_0 \leq \beta \leq \theta_0 + \Delta}
            \Big|\frac{\partial f}{\partial \theta} (\j\omega, \theta=\beta) \Big|}.
            \label{eq:line_imaginary_axis}
        \end{equation}
        Steps smaller than $\Delta$ retain stability if
        \eqref{eq:line_imaginary_axis} holds $\forall \omega \in \mathbb{R}_0^+$.
        Thus, a valid step limit can be obtained by finding the minimum of
        \eqref{eq:line_imaginary_axis} with respect to $\omega$ (the worst-case scenario),
        resulting in \eqref{eq:line_sufficient}.
        In deriving \eqref{eq:line_imaginary_axis}, the maximum is assumed to be different from zero.
        If it equals zero,
        then \eqref{eq:fundamental2} implies that $f$ is locally independent of $\theta$ and that
        $\Delta$ can be further increased.
        Hence, \eqref{eq:line_imaginary_axis} holds and the proof is concluded.
        \qed
    \end{pf}

    \begin{rem}
        Theorem \ref{thm:line_sufficient} determines
        a non-maximal stability equivalence segment along the line \eqref{eq:line_parametrization}.
        Its computational complexity is independent of $n$, the dimension of $\bm{\tau}$,
        as only the scalar $\overline{\Delta}$ is computed regardless of $n$.
        \label{rem:complexity_line_sufficient}
    \end{rem}

    \begin{rem}
        \label{rem:circularity}
        The maximal step size $\overline{\Delta}$ appears on both sides of \eqref{eq:line_sufficient},
        making the expression circular.
        However, the LHS of \eqref{eq:line_sufficient} is strictly increasing, while the RHS is non-increasing.
        Thus, a valid value of the (not necessarily maximal) step size $\Delta$ can be found by
        bisection up to a certain tolerance threshold.
        Any conservatism introduced at this point is overcome by iterating the method,
        as shown in section \ref{section:line_nsc}.
        Lastly, for retarded TDS the RHS of \eqref{eq:line_sufficient} can be substituted by a conservative form,
        independent of $\overline{\Delta}$ (hence removing circularity), as discussed below.
    \end{rem}

    \subsection*{Application to retarded TDS}
    \label{section:line_retarded}
    Although applicable to a wide class of linear systems,
    the proposed method is particularly simple in case of retarded TDS, which characteristic function is given by
    \begin{equation}
        f(s, \bm{\tau}) = s^m + \sum_{i=1}^{n} P_i(s)e^{-s\tau_i},
        \label{eq:retarded_characteristic}
    \end{equation}
    where $P_i(s)$ are polynomials with $\deg P_i(s) < m$.
    Plugging \eqref{eq:line_parametrization} into \eqref{eq:retarded_characteristic} yields
    \begin{equation}
        f(s, \theta) = s^m + \sum_{i=1}^n f_i(s)e^{-s \theta a_i},
        \label{eq:retarded_characteristic3}
    \end{equation}
    where $a_i$ are real scalars and $f_i(s)$ are complex functions,
    independent of $\theta$, that can easily be computed from \eqref{eq:retarded_characteristic}.
    This result is important because of the convenient form of \eqref{eq:retarded_characteristic3},
    which however is not limited to retarded TDS.
    Namely, to implement the general form \eqref{eq:line_sufficient}, evaluation of $|f(\j\omega, \theta_0)|$
    and $\max_{\theta_0 \leq \beta \leq \theta_0 + \Delta}
    |\frac{\partial f}{\partial \theta} (\j\omega, \beta) |$ is required.
    The former expression is directly evaluated from \eqref{eq:retarded_characteristic3}.
    For the latter, observe that
    \begin{multline}
        \max_{\theta_0 \leq \beta \leq \theta_0 + \Delta}
        \Big|\frac{\partial f}{\partial \theta} (\j\omega, \theta = \beta) \Big| \leq
        \sum_{i=1}^n \omega \big| a_i f_i(\j\omega)|,
        \label{eq:retarded_simplification}
    \end{multline}
    which yields an elegant expression, albeit conservative.
    Hence, the following corollary to Theorem \ref{thm:line_sufficient} is formulated.
    \begin{cor}
        Let $f$ be defined as in \eqref{eq:retarded_characteristic3}.
        Let $\theta_0 \geq 0$ be an initial value such that
        $f(\j\omega, \bm{\tau}(\theta_0)) \neq 0, \forall \omega \in \mathbb{R}$.
        Let $\bm{\tau}(\theta)$ be defined as in \eqref{eq:line_parametrization}. Then,
        \begin{equation}
            NU_f(\bm{\tau}(\theta_0)) = NU_f(\bm{\tau}(\theta_0 + \Delta))
        \end{equation}
        holds if
        \begin{equation}
            \Delta < \min_{\omega \in \mathbb{R}_0^+}
            \frac{|f(\j\omega, \theta)|}{\sum_{i=1}^n \omega |a_i f_i(\j\omega)|}.
            \label{eq:algorithm_retarded}
        \end{equation}
        \label{cor:line_retarded}
    \end{cor}

    This corollary presents a convenient alternative to Theorem \ref{thm:line_sufficient},
    since \eqref{eq:algorithm_retarded} bypasses the circularity of \eqref{eq:line_sufficient}.

    \begin{rem}
        The RHS of \eqref{eq:algorithm_retarded} contains a frequency sweep over $\omega \in \mathbb{R}_0^+$.
        However, in case of retarded TDS,
        the sweep can be confined to a finite interval as pointed out in \cite[Proposition 1.12]{michiels2014}.
        \label{rem:finite_sweep}
    \end{rem}

    \subsection{Stability limit}
    \label{section:line_nsc}
    Under assumptions of Theorem \ref{thm:line_sufficient}, by applying \eqref{eq:line_sufficient},
    one can obtain a stability equivalence interval defined by the endpoint
    \begin{equation}
        \theta_1 = \theta_0 + \Delta.
    \end{equation}
    The method can now be applied again, taking previously obtained value $\theta_1$ as the new starting point.
    Formally, the method can be iterated a certain number of times
    \begin{equation}
        \theta_{k+1} = \theta_k + \Delta_k, \quad \forall k \in \mathbb{N}_0.
        \label{eq:line_algorithm}
    \end{equation}
    Such an iterative application of the method converges to the stability boundary,
    since the resulting sequence $\theta_k$
    exactly converges to $\theta_{lim}$,
    as defined in \eqref{eq:theta_lim},
    provided $\theta_{lim}$ exists.
    If $\theta_{lim}$ does not exist, the sequence $\theta_k$ diverges.
    The aforementioned claim is formalized and proven in the following lemma and theorem.
    \begin{lem}
        Let the hypotheses of Theorem \ref{thm:line_sufficient} be satisfied.
        Let $\theta_{lim}$ be defined in \eqref{eq:theta_lim}
        and let a sequence $\theta_k$ be obtained by \eqref{eq:line_algorithm},
        with increments $\Delta_k = \eta \overline{\Delta}(\theta_k)$,
         $\eta\in(0,1)$ and $\overline{\Delta}(\theta_k)$
        computed according to \eqref{eq:line_sufficient} in each iteration.
        Then, the following statements hold:
        \begin{enumerate}[label=(C\arabic*)]
            \item \label{item:statement1} If $ \theta_{lim}$ exists,
                  then $\theta_k < \theta_{lim}, \forall k \in \mathbb{N}_0$.
            \item \label{item:statement2} If $\displaystyle \lim_{k \rightarrow \infty} \theta_k$ exists,
                  then $\displaystyle \lim_{k \rightarrow \infty} \theta_k = \theta_{lim}$.
        \end{enumerate}
        \label{lem:merged}
    \end{lem}
    \begin{pf}
        Claim \ref{item:statement1} is a direct consequence of Theorem \ref{thm:line_sufficient}, since $0 < \eta < 1$.
       Claim \ref{item:statement2} can be proven by contradiction.
        Assume that $\lim_{k \rightarrow \infty} \theta_k$ converges to some $ \theta_{\#} < \theta_{lim}$.
        As a consequence of \ref{item:statement1}, such $\theta_{\#}$ must be smaller than $\theta_{lim}$.
        The bare existence of a convergence limit implies that values $\Delta_k$
        get arbitrary small as $k \rightarrow \infty$.
        This, combined with \eqref{eq:line_sufficient} implies that the value of
        \begin{equation}
            \min_\omega|f(\j\omega, \theta_k)|
            \label{eq:proof}
        \end{equation} becomes arbitrary small as $k \rightarrow \infty$ and $\theta \rightarrow \theta_{\#}$.
        However, it is not possible that \eqref{eq:proof} becomes arbitrarily small near $\theta_{\#}$ because:
        \begin{enumerate}
            \item Function $|f(\j\omega, \theta)|$ is continuous with regards to $\theta$.
            \item {By definition \eqref{eq:theta_lim}, $\theta_{lim}$ is the smallest value of
            $\theta \in [\theta_0, \infty)$ for which $\exists \omega \in \mathbb{R}$ such that
                \begin{equation}
                    |f(\j\omega, \theta)|=0.
                \end{equation}
            }
        \end{enumerate}
        Thus, $\exists \alpha \in \mathbb{R}^+$ and $\exists \varepsilon \in \mathbb{R}^+$ such that
        \begin{equation}
            \min_\omega|f(\j\omega, \theta)| >
                        \alpha, \forall \theta \in (\theta_{\#}-\varepsilon, \theta_{\#}+\varepsilon),
        \end{equation}
        contradicting the assumption that $\theta_k \rightarrow \theta_{\#}$.
        In other words, values of $\Delta_k$ cannot be arbitrarily small
        in the neighborhood of any $\theta_\# < \theta_{lim}$. \qed
    \end{pf}

    \begin{thm}
        Let the hypotheses of Theorem \ref{thm:line_sufficient} be satisfied.
        Let $\theta_{lim}$ be defined in \eqref{eq:theta_lim}
        and let a sequence $\theta_k$ be obtained by \eqref{eq:line_algorithm},
        with increments $\Delta_k = \eta \overline{\Delta}(\theta_k)$, $0<\eta<1$,
        and $\overline{\Delta}(\theta_k)$ fulfilling \eqref{eq:line_sufficient}.
        If $\theta_{lim}$ exists, then $\theta_{k}$ converges to $\theta_{lim}$.
        Otherwise, $\theta_{k}$ diverges.
        \label{thm:line_nsc_convergence}
    \end{thm}
    \begin{pf}
        Assume that $\theta_{lim}$ exists.
        From \eqref{eq:line_sufficient} and since $\eta > 0$, the sequence $\theta_k$ is strictly increasing.
        From Lemma \ref{lem:merged}, the sequence $\theta_k$ will never overshoot $\theta_{lim}$.
        Hence, the sequence $\theta_k$ must converge to a value in the interval $[\theta_0, \theta_{lim}]$.
        From Lemma \ref{lem:merged}, the only possible value of convergence in the given interval is $\theta_{lim}$.

        On the other hand, assume that $\theta_{lim}$ does not exist.
        Similarly to Lemma \ref{lem:merged}, the convergence of an increasing sequence $\theta_k$ would
        imply that the values of $\min_\omega|f(\j\omega, \theta_k)|$ get arbitrary small as $k \rightarrow \infty$.
        This is not possible because the non-existence of $\theta_{lim}$ implies that $\exists \alpha > 0$ such that
        \begin{equation}
            \min_\omega|f(\j\omega, \theta_k)| > \alpha, \;\; \forall \theta > \theta_0.
        \end{equation}
        Thus, the steps $\Delta_k$ cannot become arbitrarily small, concluding the proof.\qed
    \end{pf}
    \begin{algorithm}
        \hrulefill
        \begin{algorithmic}[0]
            \Require $\delta > 0, \Theta > 0, \theta_0 \in [0, \Theta), \eta \in (0, 1)$
            \State $\theta_k := \theta_0$
            \State $\overline{\Delta}(\theta_k) := \infty$
            \While{$\eta \overline{\Delta}(\theta_k) > \delta \text{ and } \theta_k < \Theta$}
                \State $\overline{\Delta}(\theta_k) :=  \displaystyle \min_{\omega}\frac{|f(\j\omega, \theta_k)|} 
                    {\max_{\theta_k \leq \beta \leq \theta_k + \overline{\Delta}(\theta_k)}
                    \Big|\frac{\partial f}{\partial \theta} (\j\omega, \theta=\beta) \Big|}$
                \State $\theta_k := \theta_k + \eta \overline{\Delta}(\theta_k)$
                \State $k := k + 1$
            \EndWhile
            \State \Return $\theta_{lim} := \theta_k$
        \end{algorithmic}
        \caption{Approximate computation of $\theta_{lim}$}
        \label{alg:line}
        \hrulefill
    \end{algorithm}
    \paragraph*{Implementation issues.}The procedure for approximate evaluation of
    $\theta_{lim}$ is presented in Algorithm~\ref{alg:line}.
    Numerical implementation of the algorithm
    introduces issues related to the floating point representation of small and large numbers.
    If $\theta_k$ is convergent,
    then the steps $\overline{\Delta}(\theta_k)$ converge towards zero as $\theta_k$ iteratively increases.
    Since the computer precision is finite,
    a termination criterion is introduced when $\eta \overline{\Delta}(\theta_k)$
    becomes smaller than a prescribed value $\delta$. On the other hand,
    since the algorithm cannot be run indefinitely,
    another termination criterion is introduced when $\theta_k$ becomes larger than a prescribed value $\Theta$. 
    Hence, if the algorithm returns a value greater than or equal to $\Theta$,
    it indicates that either the sequence is divergent,
    or that the stability limit $\theta_{lim}$ is beyond the considered searching domain.
    Increasing $\Theta$ mitigates this problem to a certain extent at the cost of an increased number of iterations.
    Finally, \eqref{eq:line_sufficient} depends on finding the global minimum of a function.
    If the minimum is overestimated due to numerical issues related to the
    finite precision of floating point arithmetics, an accidental jump,
    $\overline{\Delta}(\theta_k)$, of $\theta$ beyond the true stability limit, $\theta_{lim} < \Theta$,
    may occur, leading to a wrong evaluation of $\theta_{lim}$.
    Thus, care must be taken, when performing the necessary global optimizations, to avoid such accidental jumps.
    This is precisely the reason why the scaling factor $\eta \in (0, 1)$ is introduced
    in Lemma \ref{lem:merged}, and Theorem \ref{thm:line_nsc_convergence}.

    \begin{rem}
        Theorem \ref{thm:line_nsc_convergence} determines
        the maximal stability equivalence segment along the line \eqref{eq:line_parametrization}.
        Its computational complexity is independent of $n$, the dimension of $\bm{\tau}$.
        \label{rem:complexity_line_nsc}
    \end{rem}

    \begin{rem}
        Instead of extending stability along a line as in \eqref{eq:line_parametrization},
        any smooth curve parametrized by a scalar $\theta$,
        provided that $\bm{\tau}(\theta_0 = 0) = \bm{\tau^0}$, could have been chosen.
        For example, one might analyze stability along an arc of an \mbox{n-sphere}.
        \label{rem:LineCurve}
    \end{rem}

    \begin{exmp}
            Consider a system modeled by
            \begin{equation}
                \dot{x}(t) = -x(t-\tau_1) -\int_{-\tau_2}^{0} e^{k\alpha} x(t+\alpha) d\alpha.
                \label{eq:example_3d_ss_model}
            \end{equation}
            Its stability is investigated with respect to $\tau_1$, $\tau_2$, and $k$.
        \label{example:example_3d}
    \end{exmp}

\begin{figure}
        \begin{center}
                \centering
                \includegraphics[scale=1.2]{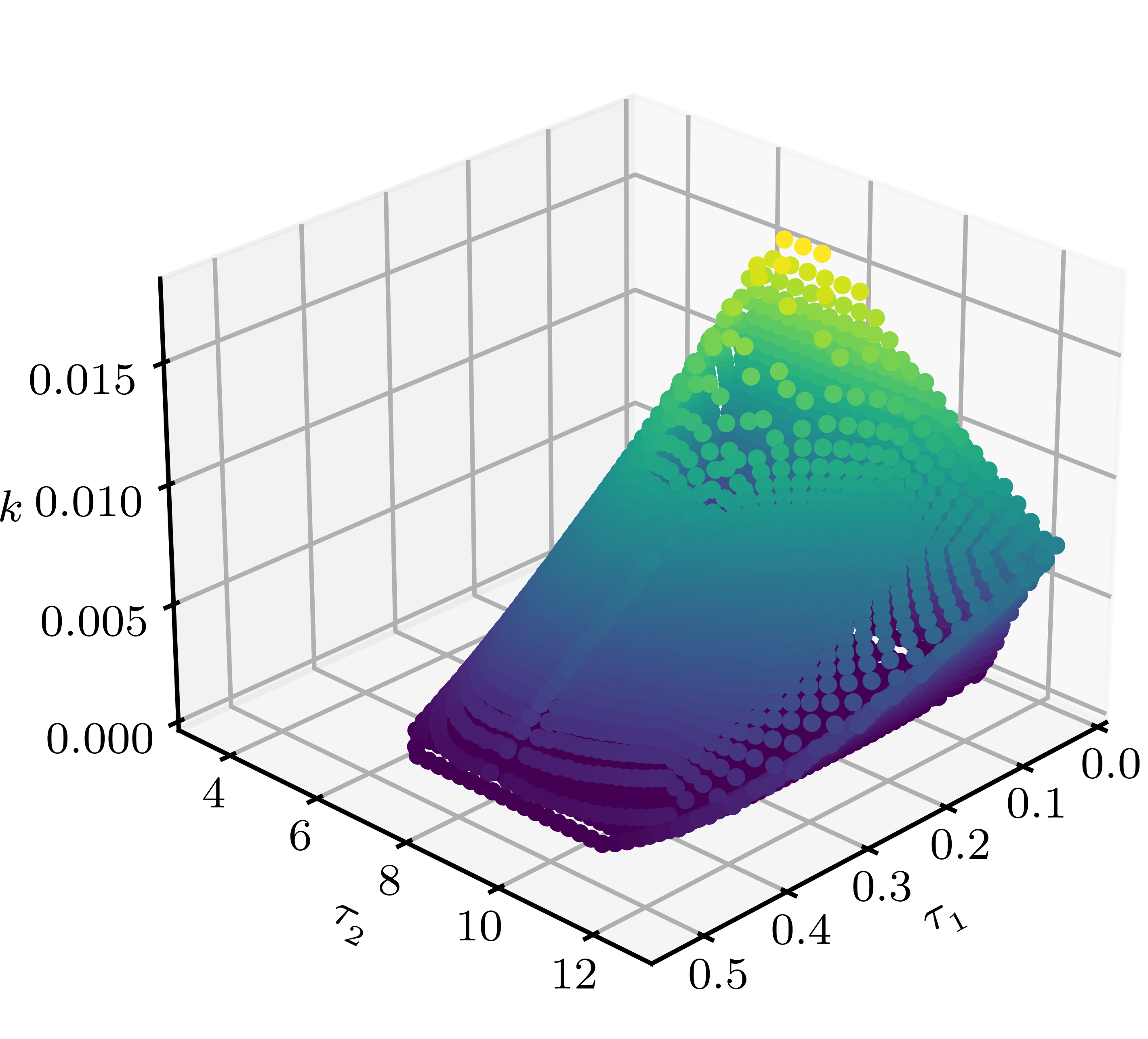}
            \caption{Stability analysis of Example \ref{example:example_3d},
            computed from the starting point $(\tau_1, \tau_2, k) = (0.250, 8.000, 0.003)$}
            \label{fig:example_3d}
        \end{center}
    \end{figure}

        Stability of \eqref{eq:example_3d_ss_model} can be reduced to the analysis of
        \begin{equation*}
            f(s,\tau_1, \tau_2, k) = s^2 + s(k+e^{-s\tau_1}) + k e^{-s\tau_1} + 1 - e^{-\tau_2(k+s)},
        \end{equation*}
        which fulfills \ref{hypothesis1} and \ref{hypothesis2}.
        Algorithm \ref{alg:line} is applied to a manually chosen starting point
        $(\tau_1, \tau_2, k) = (0.250, 8.000, 0.003)$.
        The endpoints of obtained stability equivalence rays are plotted in Fig. \ref{fig:example_3d}.
        \qed

\begin{figure}
        \begin{center}
            \includegraphics[scale=0.85]{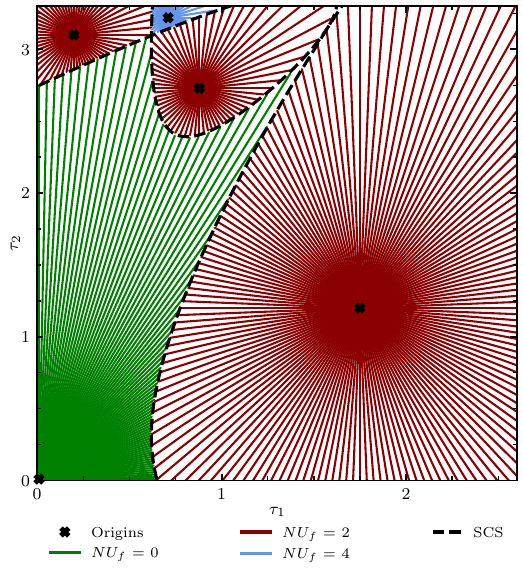}
            \caption{Stability analysis of Example \ref{example:example_retarded}}
            \label{fig:example_retarded_line}
        \end{center}
    \end{figure}

    \begin{exmp}
        Consider a system with a characteristic function given by
        \begin{equation}
            f(s, \bm{\tau}) = s^2 + 2se^{-s\tau_1} + e^{-s\tau_2}.
        \end{equation}
        Its stability is investigated with respect to $\bm{\tau} = [\tau_1, \tau_2]$.
        \label{example:example_retarded}
    \end{exmp}
    Since the system is retarded, the simplified version of the algorithm
    (using Corollary \ref{cor:line_retarded}) is applied.
    The algorithm is initialized at five different points,
    for which the number of unstable poles has been determined using Cauchy's argument principle.
    The results are displayed in Fig. \ref{fig:example_retarded_line} and compared to
    the stability crossing set (SCS) obtained by \cite{gu2005} for verification purposes.

    Although applying the algorithm to obtain a
    plethora of rays gives a good sketch of the stability equivalence region,
    the result does not guarantee stability equivalence in a dense set of $(\tau_1, \tau_2)$.
    This shortcoming is overcome in section \ref{section:region} by analyzing stability inside a region.\qed

    \section{Stability equivalence within a region}
    \label{section:region}
    In this section, a solution to problem \ref{problem2} is obtained.
    \subsection{Sufficient condition}
    \begin{thm}
        Let $f$ satisfy \ref{hypothesis1} and \ref{hypothesis2}.
        Let $\bm{\tau^0} \in \mathcal{T}$ be any parameter point satisfying
        $f(\j\omega, \bm{\tau^0}) \neq 0, \forall \omega \in \mathbb{R}$. Let $p$ and $q$ satisfy
        \begin{equation}
            \frac{1}{p} + \frac{1}{q} = 1, \quad 1 \leq p, q \leq \infty.
            \label{eq:pq}
        \end{equation}
        Then,
        $$NU_f(\bm{\tau^0}) = NU_f(\bm{\tau}^0 + \mathbf{v})$$
        holds for every $\mathbf{v}$ such that $\|\mathbf{v}\|_q < \overline{\varepsilon}_{p,q}(\bm{\tau}^0)$, where
            \begin{equation}
                \overline{\varepsilon}_{p,q}(\bm{\tau}^0) =
                \min_{\omega \in \mathbb{R}_0^+} \frac{|f(\j\omega, \bm{\tau^0})|}{\displaystyle
                \max_{||\mathbf{v}||_q \le
                \overline{\varepsilon}_{p,q}(\bm{\tau}^0)}||\nabla f(\j\omega,\bm{\tau^0} + \mathbf{v})||_p}.
                \label{eq:region_final_theorem}
        \end{equation}
        \label{thm:region_sufficient}
    \end{thm}

    \vspace{-0.75cm}
    \begin{pf}
        To build towards the proof,
        it is beneficial to start by analyzing stability equivalence of two arbitrary parameter points.
        To that end, define a parameter point $\bm{\tau}$ as
        \begin{equation}
            \bm{\tau}(\mathbf{v}) = \bm{\tau^0} + \mathbf{v},
        \end{equation}
        where $\bm{\tau^0} \in \mathcal{T}$ represents a
        chosen starting point and $\mathbf{v}$ represents a change vector.
        The objective is to discuss the stability equivalence of parameter points
        $\bm{\tau^0} = \bm{\tau}(\mathbf{0})$ and $\bm{\tau}(\mathbf{v})$.
        From Rouch\'e's theorem, it is known that stability equivalence of these points is guaranteed if
        \begin{multline}
            |f(s,\bm{\tau}(\mathbf{v})) - f(s, \bm{\tau}(\mathbf{0}))| <
                            |f(s, \bm{\tau}(\mathbf{0}))|, \forall s \in \mathcal{C}.
            \label{eq:region1}
        \end{multline}
        The LHS of \eqref{eq:region1} can further be elaborated to obtain
        \begin{align*}
            &|f(s, \bm{\tau}(\mathbf{v})) - f(s, \bm{\tau}(\mathbf{0}))| =
                \Bigg|\int_{\gamma} \Big\langle \nabla f(s, \bm{\tau}(\mathbf{r})), d\mathbf{r} \Big\rangle \Bigg|= \\
            &\Bigg|\int_0^1
                \Big\langle \nabla f (s, \bm{\tau}(\mathbf{r}(\beta))), \mathbf{r'} \Big\rangle d\beta \Bigg| \leq \\
            &\int_0^1 \Bigg|
                \Big\langle \nabla f (s, \bm{\tau}(\mathbf{r}(\beta))), \mathbf{r'} \Big\rangle \Bigg| d\beta ,
            \label{eq:region2}
        \end{align*}
        where $\mathbf{r}(\beta)$ represents parameterization of curve $\gamma$ which connects the $\mathbf{0}$ vector
        with $\mathbf{v}$ for $\beta \in [0, 1]$,
        and $\mathbf{r'}$ represents the derivative of $\mathbf{r}(\beta)$ with respect to $\beta$.
        Introducing the obtained conservative bound in \eqref{eq:region1},
        and using \ref{hypothesis2} and the symmetry of $f$, implies that for every $\omega>0$
        \begin{equation*}
            \int_0^1 \Bigg| \Big\langle \nabla f (\j\omega, \bm{\tau}(\mathbf{r}(\beta))),
            \mathbf{r'} \Big\rangle \Bigg| d\beta < |f(\j\omega, \bm{\tau}(\mathbf{0}))| \;.
            \label{eq:region2.5}
        \end{equation*}
        In order to simplify notation, in the remainder of this proof $f(\j\omega, \bm{\tau}(\mathbf{v}))$ and
        $\nabla f(\j\omega, \bm{\tau}(\beta \mathbf{v}))$
        are denoted as $f(\mathbf{v})$ and $\nabla f(\beta \mathbf{v})$, respectively.
        By defining the curve $\gamma$ as $\mathbf{r}(\beta) = \beta\mathbf{v}$ and applying H\"older's inequality,
        \begin{equation}
            \int_0^1 \Bigg|\Big\langle \nabla f (\beta\mathbf{v}), \mathbf{v} \Big\rangle
                \Bigg|d\beta \leq \int_0^1 ||\nabla f(\beta\mathbf{v})||_p ||\mathbf{v}||_q\;d\beta.
            \label{eq:region3}
        \end{equation}
        The results presented so far guarantee stability equivalence for a specific change vector $\mathbf{v}$.

        Choosing arbitrary positive $\varepsilon_{p,q}$,
        one may notice that for any $\mathbf{v}$ which satisfies $||\mathbf{v}||_q \le \varepsilon_{p,q}$,
        it is possible to substitute \eqref{eq:region3} with a more conservative expression
        \begin{equation*}
            \int_0^1 ||\nabla f(\beta\mathbf{v})||_p\;||\mathbf{v}||_q\;d\beta \; \leq \;
                \max_{||\mathbf{v}||_q<\varepsilon_{p,q}}||\nabla f(\mathbf{v})||_p \; \varepsilon_{p,q} \;,
            \label{eq:region4}
        \end{equation*}
        derived from the fact that the integral of a positive quantity is always less
        or equal than the product of the maximum of the integrand by the length of the integration interval.
        Finally, the upper bound on $\varepsilon_{p,q}$, denoted as $\overline{\varepsilon}_{p,q}(\bm{\tau}^0)$,
        defining the  permissible stability equivalence region, is obtained as in \eqref{eq:region_final_theorem},
        concluding the proof. \qed
    \end{pf}

    \begin{rem}
        Theorem \ref{thm:region_sufficient} determines a non-maximal
        stability equivalence region surrounding a given parametric point.
        Its computational complexity is independent of $n$, the dimension of $\bm{\tau}$,
        as only the scalar $\overline{\varepsilon}_{p,q}$ is computed regardless of $n$.
        \label{rem:complexity_region_sufficient}
    \end{rem}

    \begin{rem}
            \label{rem:circularity_region}
            The inequality \eqref{eq:region_final_theorem} is circular,
            since $\overline{\varepsilon}_{p,q}(\bm{\tau}^0)$ appears on both sides.
            Similarly to \eqref{eq:line_sufficient},
            the monotonicity of the involved expressions allows
            finding a valid value of $\overline{\varepsilon}_{p,q}$ by bisection.
            Moreover, specific system types (such as \eqref{eq:retarded_characteristic}) allow direct
            evaluation of the RHS,
            removing the circularity, as discussed below.
    \end{rem}

    \subsection*{Application to retarded TDS}
    \label{section:region_retarded}
    Analogously to the line-based version of the method,
    the convenient form of retarded TDS characteristic function
    given by \eqref{eq:retarded_characteristic} can be utilized to further simplify \eqref{eq:region_final_theorem}.
    To determine $||\nabla f (\j\omega, \bm{\tau})||_p$ in \eqref{eq:region_final_theorem},
    it is beneficial to first evaluate partial derivatives of
    $f$ with regards to each component $\tau_i$.  Assuming $\omega \geq 0$, observe that
    \begin{equation}
        \bigg| \frac{\partial f}{\partial \tau_i} (\j\omega, \bm{\tau}) \bigg| =
            \bigg| -\j\omega P_i(\j\omega)e^{-\j\omega\tau_i} \bigg| = \omega \bigg| P_i(\j\omega) \bigg|
    \end{equation}
    which allows expressing the norm of $\nabla f$ as
    \begin{equation}
        ||\nabla f (\j\omega, \bm{\tau})||_p =
                \Bigg(\sum_{i=1}^n \Big( \omega \big|P_i(\j\omega)\big|\Big)^p \Bigg)^{\frac{1}{p}}
    \end{equation}
    which does not depend on $\bm{\tau}$ and
    thus removes the circularity from \eqref{eq:region_final_theorem}.
    \begin{cor} \label{cor:final}
        Let $f$ be defined as in \eqref{eq:retarded_characteristic}.
        Let $\bm{\tau^0} \in \mathcal{T}$ be any parameter point satisfying
        $f(\j\omega, \bm{\tau^0}) \neq 0, \forall \omega \in \mathbb{R}$.
        Let $p$ and $q$ satisfy \eqref{eq:pq}.
        Then, $$NU_f(\bm{\tau^0}) = NU_f(\bm{\tau}^0 + \mathbf{v}), $$ holds if
        \begin{equation}
            ||\mathbf{v}||_q < \min_{\omega \in \mathbb{R}_0^+} \frac{|f(\j\omega, \bm{\tau^0})|}
                    {\Big(\sum_{i=1}^n \big( \omega|P_i(\j\omega)|\big)^p \Big)^{\frac{1}{p}}}.
            \label{eq:region_retarded_final}
        \end{equation}
    \end{cor}

\begin{figure}
        \begin{center}
            \includegraphics[scale=0.9]{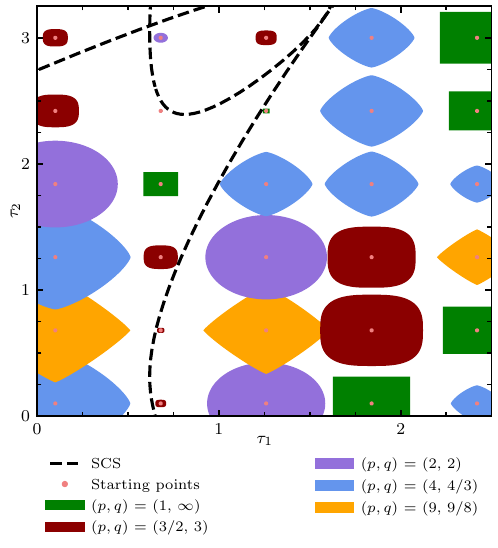}
            \caption{Results of applying Corollary \ref{cor:final}
                to various parametric points in Example \ref{example:example_retarded}.}
            \label{fig:example_retarded_random_regions}
        \end{center}
    \end{figure}

    The application of \eqref{eq:region_final_theorem} is analogous to
    performing a single step of the line version algorithm.
    Likewise, Remark \ref{rem:finite_sweep} is applicable to \eqref{eq:region_retarded_final} as well.
    Fig. \ref{fig:example_retarded_random_regions} shows the results of applying
    Corollary \ref{cor:final} to Example \ref{example:example_retarded},
    with different shapes corresponding to different combinations of $(p, q)$ and different starting points.
    The number of unstable poles is equivalent for all the points inside each individual region.

    \subsection{Maximal stability equivalence region} \label{section:region_nsc}
    Analogously to the line-based version,
    an iterative method for finding the maximal surrounding parametric region
    is established,
    in which the number of unstable poles is invariant.
    First, choose $p$ and $q$ satisfying \eqref{eq:pq}, and $\eta \in (0, 1)$.
    Choose a starting point $\bm{\tau}^0$ and define a set $\mathcal{S}_0$ as
    \begin{equation}
        \mathcal{S}_0 = \{\bm{\tau^0}\}.
        \label{eq:region_s0}
    \end{equation}
    Construct a monotonously growing sequence of sets
    \begin{equation}
        \mathcal{S}_{k+1} = \mathcal{S}_k \cup \bigcup_{\bm{\tau} \in
            \partial \mathcal{S}_k} \mathcal{W}_{\eta}(\bm{\tau}), \quad \forall k \in \mathbb{N}_0 \;,
        \label{eq:region_iterative}
    \end{equation}
    where
    \begin{equation}
        \mathcal{W}_{\eta}(\bm{\tau}) = \left\{(\bm{\tau} + \mathbf{v}) \in \mathcal{T} \;
            \Big| \; ||\mathbf{v}||_q \leq \eta \overline{\varepsilon}_{p,q}(\bm{\tau})\right\} \;,
        \label{eq:w}
    \end{equation}
    with $\overline{\varepsilon}_{p,q}(\bm{\tau})$ defined in \eqref{eq:region_final_theorem}.
    It is now established that $\mathcal{S}_k$ converges to $\mathcal{M}_f(\bm{\tau^0})$.
    \begin{thm}
        Let $f$ satisfy \ref{hypothesis1} and \ref{hypothesis2}.
        Let $p$ and $q$ satisfy \eqref{eq:pq}. Let $\bm{\tau^0} \in \mathcal{T}$ be any parameter point
        satisfying $f(\j\omega, \bm{\tau^0}) \neq 0, \forall \omega \in \mathbb{R}$.
        Define $\mathcal{S}_k$, $k\ge0$ as in \eqref{eq:region_s0} and \eqref{eq:region_iterative}. Then,
        \begin{equation}
            \limsup_{k \rightarrow \infty} \mathcal{S}_k = \mathcal{M}_f(\bm{\tau^0}).
        \end{equation}
        \label{thm:region_nsc}
    \end{thm}
    \vspace{-0.75cm}
    \begin{pf}
        Choose any point $\bm{\tau^*} \in \mathcal{M}_f(\bm{\tau^0})$.
        By definition of $\mathcal{M}_f(\bm{\tau^0})$, there exists a path $\mathcal{P}$
        defined by a continuous bijective function
        $g : [0,1] \rightarrow \mathcal{P} \subset \interior\big(\mathcal{M}_f(\bm{\tau^0})\big)$
        such that $g(0) = \bm{\tau^0}$ and $g(1) = \bm{\tau^*}$.
        Define the sequence
        \begin{equation}
            m_k = \max \left\{x \in [0,1] \; \Big| \; g(x) \in \mathcal{S}_k\right\}.
            \label{eq:m_k}
        \end{equation}
        For any fixed $k$, the set $\mathcal{S}_k$ is closed and bounded, and therefore compact.
        Consequently, the maximum in \eqref{eq:m_k} is well-defined.
        Define the sequence $\bm{\tau^k} = g(m_k)$, which represents the farthest point along the path $\mathcal{P}$
        (referenced from $\bm{\tau^0}$) such that $\bm{\tau^k} \in \mathcal{S}_k$ at iteration $k$.
        There are two possible scenarios:
        \begin{enumerate}
            \item \label{scenario1}
                $\bm{\tau^{k-1}} \neq \bm{\tau^*}$, implying $\bm{\tau^k} \in \partial \mathcal{S}_k$.
                  In this scenario, $\bm{\tau^k}$ is one of the points on which
                  \eqref{eq:region_iterative} is evaluated at iteration $k$.
            \item \label{scenario2}
                $\bm{\tau^{k-1}} = \bm{\tau^*}$,
                implying that the endpoint $\bm{\tau^*}$ has already been reached.
        \end{enumerate}

        Let us further analyze scenario (\ref{scenario1}).
        Since $\bm{\tau^k} \in \interior\big(\mathcal{M}_f(\bm{\tau^0})\big)$,
        it holds that $|f(\j\omega, \bm{\tau^k})| > 0, \forall k \in \mathbb{N}_0, \forall \omega \geq 0$,
        further implying that the resulting $\varepsilon_{p,q}$ from
        \eqref{eq:region_final_theorem} is strictly positive $\forall k \in \mathbb{N}_0$.
        Consequently, either $m_k = 1$, or $m_k < m_{k+1}$, meaning that $\bm{\tau^k}$ gets
        strictly closer to $\bm{\tau}^*$ along $\mathcal{P}$ at
        each successive iteration unless $\bm{\tau^k} = \bm{\tau^*}$.
        Thus, $\exists k_0$ such that $\bm{\tau^*} \in \mathcal{S}_{k}, \forall k \geq k_0$.
        Since the same reasoning can be applied to any chosen point $\bm{\tau^*} \in \mathcal{M}_f(\bm{\tau^0})$,
        the proof is concluded. \qed
    \end{pf}

    \begin{rem}
        \label{rem:region_set_construction}
        In practice, the sequence $\mathcal{S}_k$ is constructed on a finite set of
        sampled points belonging to $\partial \mathcal{S}_k$,
        rather than an infinite set of points as in \eqref{eq:region_iterative}.
    \end{rem}

    \begin{exmp} \label{ex:new}
        Consider a distributed delay system modeled by
            \begin{equation}
                \dot{x}(t) = -\int_{-\tau}^{0} e^{k\alpha} x(t+\alpha) d\alpha.
                \label{eq:example_new}
            \end{equation}
            Its stability is investigated with respect to $\tau$ and $k$.
    \end{exmp}
    Stability of \eqref{eq:example_new} can be reduced to the analysis of
    \begin{equation}
        f(s,\tau,k) = s^2+sk+1-e^{-\tau(s+k)},
    \end{equation}
    which fulfills \ref{hypothesis1} and \ref{hypothesis2}.
    \begin{figure}
        \begin{center}
            \includegraphics[scale=0.8]{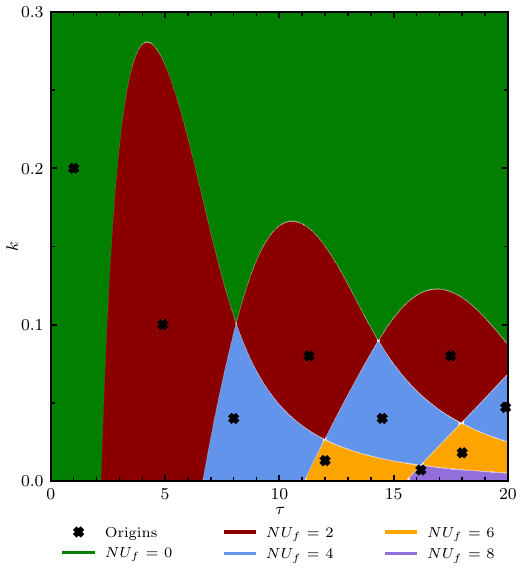}
            \caption{Results of iteratively applying \eqref{eq:region_final_theorem}
            to Example \ref{ex:new}}
            \label{fig:example_distributed_region}
        \end{center}
    \end{figure}
    Fig. \ref{fig:example_distributed_region} shows the results of applying \eqref{eq:region_iterative} to
    ten different starting points, for which the number of unstable poles
    are determined using Cauchy's argument principle.
    The obtained stability regions are similar regardless which starting point is chosen
    in the interior of the represented regions\footnote{If a starting point is on the boundary of two regions,
    then condition $f(\j\omega, \bm{\tau^0}) \neq 0$ is not satisfied and Theorems \ref{thm:line_sufficient},
    \ref{thm:line_nsc_convergence}, \ref{thm:region_sufficient} and \ref{thm:region_nsc} cannot be applied.}.
    The algorithm reached boundaries of the search space in the positive direction (up and right).
    This could be an indication that the system is unstable independently of $\tau$ and $k$
    in the areas of the parametric space lying beyond the search region (in these directions).
    \qed

\begin{figure}
        \begin{center}
            \includegraphics[scale=0.82]{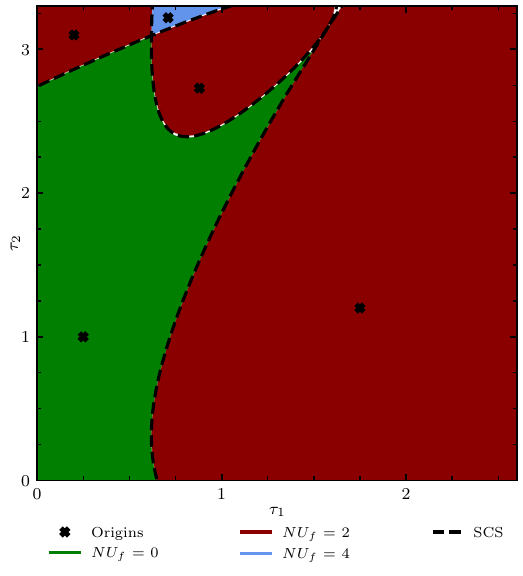}
            \caption{Results of iteratively applying \eqref{eq:region_retarded_final}
                     to Example \ref{example:example_retarded}}
            \label{fig:example_retarded_region}
        \end{center}
    \end{figure}

    Additionally, Fig. \ref{fig:example_retarded_region} shows the results of iterative applications of
    Corollary \ref{cor:final} to Example \ref{example:example_retarded}.

\begin{figure}
        \begin{center}
            \includegraphics[scale=0.82]{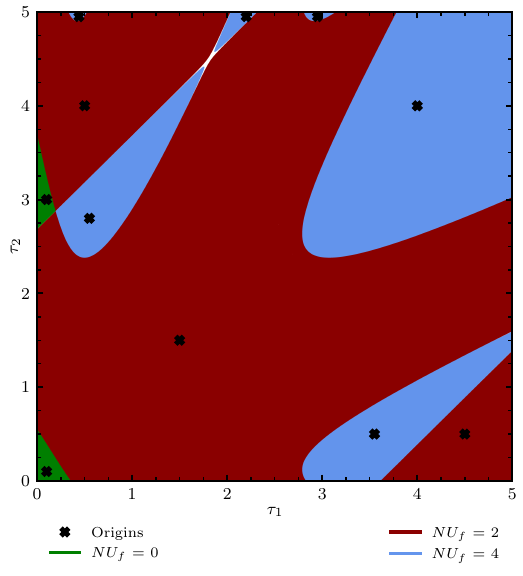}
            \caption{Results of iteratively applying \eqref{eq:region_retarded_final}
                     to Example \ref{example:degenerate}}
            \label{fig:example_degenerate}
        \end{center}
    \end{figure}

    \begin{exmp} \label{example:degenerate}
        Consider a system with a characteristic function given by
        \begin{equation}
            f(s, \bm{\tau}) = (s^2+1) + (s+2) e^{-\tau_1 s} + \sqrt{5}e^{-\tau_2 s}
            \label{eq:retarded_degenerate}
        \end{equation}
        Its stability is investigated with respect to $\bm{\tau} = [\tau_1, \tau_2]$.
    \end{exmp}
    This example belongs to a class of systems considered to be degenerated in \cite{gu2005},
    thus requiring special considerations in their work.
    Contrary to this, the approach proposed in this paper
    handles this example straightforwardly, with no need for special considerations of any kind.
    Fig. \ref{fig:example_degenerate} shows the results of applying
    Corollary \ref{cor:final} to \eqref{eq:retarded_degenerate} starting from 11 different points
    in the $\tau_1$ versus $\tau_2$ plane.
    Two stability regions exist in the searching domain. \qed

    \begin{rem}
        \label{remark:stability_radius}
        The region-based method can be compared to methods tackling robust stability and robust control, such as
        \cite{gu2007, knospe2006, hinrichsen1986, kressner2006}.
        However, the problems being solved are slightly different.
        When dealing with robust stability problems,
        the aim is often to find the stability radius of a given parametric point
        (minimal distance from the given parametric point to the stability crossing set).
        On the other hand, the methodology proposed in this paper aims at
        finding the entire stability equivalence region surrounding a given parametric point.
    \end{rem}

   \section{Conclusions and discussions}
    \label{section:conclusion}
    This paper presents a new methodology for analyzing stability of a large class of linear TDS,
    including retarded and distributed delays.
    The presented methods allow finding the maximal line segment and the maximal region in which
    the number of unstable poles is invariant.
    The primary comparative advantage of the proposed methodology is that it can be applied in a uniform manner to
    a wide class of problems: the only conditions are captured by hypotheses (H1) and (H2).
    It is worth emphasizing, however, that the time complexity of the line-based Algorithm~\ref{alg:line},
    which allows to reach the boundary of the stability domain in a prescribed direction,
    is independent of the dimension of the parametric vector $\bm{\tau}$.
    The method developed in this paper has further been extended to irrational systems in \cite{turkulov2023}.
    The proposed method is not able to address neutral-type systems since they violate hypothesis (H2).
    Weakening (H2), and consequently extending the results to neutral-type systems,
    is an interesting perspective of this work.
    Further perspectives are linked to establishing a computationally efficient algorithm for finding the whole or
    even all stability regions in a prescribed $n$-dimensional space using the results of
    either the line-based stability or the region-based stability.

    \bibliographystyle{newapa}
    \bibliography{references}
\end{document}